\documentclass[conference]{IEEEtran}
\usepackage{booktabs}
\newcommand{\ra}[1]{\renewcommand{\arraystretch}{#1}}

\usepackage[caption=false,font=footnotesize]{subfig}

\usepackage{amsfonts}
\usepackage{amsmath}
\usepackage{amssymb,mathrsfs}
\usepackage{array}
\usepackage{flafter}

\usepackage{verbatim}

\usepackage{color}
\usepackage{psfrag}
\usepackage{umoline}
\usepackage{hyperref}
\usepackage{indentfirst}

\allowdisplaybreaks[1]
\makeindex             

\usepackage[T1]{fontenc}
\usepackage[utf8]{inputenc}

\usepackage{lipsum}

\usepackage[figuresright]{rotating}

\begin{document}
%
\title{A Data-Driven Approach for Modeling Stochasticity in Oil Market }

\author{\IEEEauthorblockN{Sina Aghaei}
\IEEEauthorblockA{Deparment of Industrial and \\ Systems Engineering\\
University of Southern California\\
California, Los Angeles 90007\\
Email: saghaei@usc.edu }
}


%


\maketitle

\begin{abstract}
Global oil price is an important factor in determining many economic variables in the world's economy. It is generally modeled as a stochastic process and have been studied through different techniques by comparing the historic time series of demand, supply and the price itself. However, there are many historic events where the demand or supply changes are not sufficient in explaining the price changes. In such cases, it is the expectations on the future changes of demand or supply that causes heavy and quick influences on the price. There are many parameters and variables that shape these expectations, and are usually neglected in traditional models. In this paper, we have proposed a model based on System Dynamics approach that takes into account these non-traditional factors. The validity of the proposed model is then evaluated using real and potential scenarios in which the proposed model follows the trend of the real data. 
\end{abstract}

\section{Introduction}\label{sec:intro}
It is widely accepted that the price of a commodity is determined by the interacting and opposing forces of the demand and supply in any free market \cite{abel2008macroeconomics}. Like any other market this statement is correct for the oil market, as well. However, in the oil market, the total demand and supply are not the only determinants of the price \cite{chevillon2009physical,kilian2014oil, sadorsky2006modeling, rafieisakhaei2016analysis}. There are several examples in the history of oil market where the oil price has experienced sharp jumps in such a short period of time that the existing models have failed to capture the causes and reasons \cite{chiroma2015evolutionary, shin2013prediction, he2012crude,fan2008generalized}. 

It is important to note that the types of oil production resources changes a lot both geographically and in time \cite{nehring1978giant, rafieisakhaei2017effects}. Particularly, the advances of researches in the oil-related industries not only introduce new resources in the market, but they also make the costs of the production in the conventional oil wells much lower \cite{16_macrae_2011}. This makes the old models which aggregate the total production as merely one type to be less effective in capturing the sharp changes of the price \cite{sterman1988modeling, fan2008generalized}. Moreover, some of the other models like \cite{he2012crude, salisu2013modelling,musaddiq2012modeling, narayan2007modelling, yang2002analysis} model the oil price as a stochastic process, and are successful in characterizing the oil price features including its volatility. However, these models usually do not consider the underlying factors and variables in determining the oil price. For instance, parts of added demand on the market has been responded through the growth of the production in the shale industry \cite{bartis2005oil,schmidt2003new}. Particularly, US oil drillers has increased their production by 70$ \% $ since 2008, reducing the US oil imports from the Organization of Petroleum Exporting Countries (OPEC) by 50$ \% $ \cite{2_krauss_2015}. Likewise, production of oil from the Tar sands of Canada \cite{demaison1977tar} holds another reason for the increased supply in the oil market. As reported by the International Energy Agency (IEA) and OPEC, the average production of OPEC has only increased from 29.81 mb/d to 30.52 mb/d since the last quarter of 2014 till the first quarter of 2015, whereas the global oil supply has increased from 91.89 mb/d to 94.34 mb/d, which is much more significant \cite{23_generator_2015}. 

Similar to the carbon market \cite{rafieisakhaei2017efficacy}, the nature of the supply resources are different in the oil market. Henceforth, the risk factors and dependencies of the variables that are involved in the oil production process are different for different resources. For instance, the cost of setting up a conventional oil well is usually lower in the Middle East \cite{4_the_economist_2014} whereas it is much higher in the northern parts of Russia. Moreover, the cost of setting up a shale oil well is much less than that of the conventional wells, both time and money-wise \cite{4_the_economist_2014}. However, the amount of oil that is produced from the latter is much more while the shale oil wells' production usually decreases sharply by 60-70$ \% $ after the first year \cite{5_the_economist_2014}. Therefore, the shale oil industry needs a constant rate of investment as the new oil wells need to be created with much higher rates. These show that the variables involving the supply chain on the different types of resources need to be modeled with different mathematical formulas. For instance, researchers say that the current break-even price for the American projects is around \$65-70 \cite{5_the_economist_2014}. However, the producing wells are extremely profitable. On the other hand, the conventional oil well producers can tolerate a much lower price. However, many of the OPEC countries' and Russia's budget is heavily dependent on oil and this is one of the factors that reduces the OPEC countries' tolerance of the very low prices \cite{6_kvue_2015}.

To address this issue, the proposed model considers the effect that expectation whether on the demand side or supply side could have of the oil price. Details of the model is described in the following section.  

\section{Model, Loops and Formula}\label{sec:Model, Loops and Formula}
In this section, we provide the proposed model, its important loops and some of the main formula that was used in the model.

\subsection{Utilized Numerical Methods for Integration}
\begin{figure}[ht!]
	\centering
	{\includegraphics[width=1.5in]{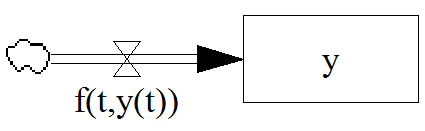}}
	\caption{Stock and flow model in Vensim.\label{fig:stock_flow} }
\end{figure}

In this subsection, we provide the mathematical basis that has been used in the model. We have used Euler method and Runge-Kutta methods to numerically solve the differential equations that result from modeling the stock variables. Let us suppose that we are given a differential equation of the form
\begin{align}
\frac{dy}{dt} &= f(t, y(t)),\label{eq:differential equation}\\
\nonumber y &= y_0;
\end{align}
where $ f $ is a Lipschitz continuous function, i.e. there exists a real number $ C>0 $ such that
\begin{align*}
|\frac{f(t_k, y(t_k))-f(t_k,y_k)}{y(t_k)-y_k}|<C,
\end{align*}
where $ y(t_k) $ is the real value of the stock variable $ y $, and $ y_k $ is the numerical approximation of this value at time step $ t_k $. This equation can be modeled in Vensim \cite{VensimP} as figure \ref{fig:stock_flow}.

\textit{Euler method:} In this method, the first order Taylor series expansion is used to approximate the value of the stock variable. In particular, at time step $ t_k $, $ y $ is approximated as:
\begin{align}
y_{k+1} = y_k+\delta tf(t_k,y_k), ~1\le k\le K
\end{align}
Therefore, the new value $ y_{k+1} $ is computed using the previous value $ y_{k} $ succeeded by moving forward using the slope given by \eqref{eq:differential equation}. It can be shown that by using this method, both global and local error decay linearly with $ \delta t $. Therefore, it is important to choose small value for $ \delta t $. One of the disadvantages of this method is that it requires explicit form of derivatives of $ f $. However, we will obtain an approximation of the derivative using the real data, which we discuss later.

\textit{Forth order Runge-Kutta method:} This method is still a one step method but utilizes the value of the variable at different points. The value $ y_k $ at time step $ t_k $ is approximated using the following set of equations:
\begin{align}
\nonumber r_1 & = \delta t f(t_k, y_k)\\
\nonumber r_2 & = \delta t f(t_k + 0.5\delta t, y_k+0.5 r_1)\\
\nonumber r_3 & = \delta t f(t_k + 0.5\delta t, y_k+0.5 r_2)\\
\nonumber r_2 & = \delta t f(t_k + \delta t, y_k+ r_3)\\
y_{k+1} &= y_k +(r_1+r_2+r_3+r_4)/6.
\end{align}
The advantage in using this method is that, it does not need explicit form of the derivative at time $ t_k $.

Now that we have provided the mathematical basis for our analysis, we proceed to provide the regression results based on this analysis, in addition to the main loops of the model.
\subsection{The Main Loop}
\begin{figure}[ht!]
	\centering
	{\includegraphics[width=3.5in]{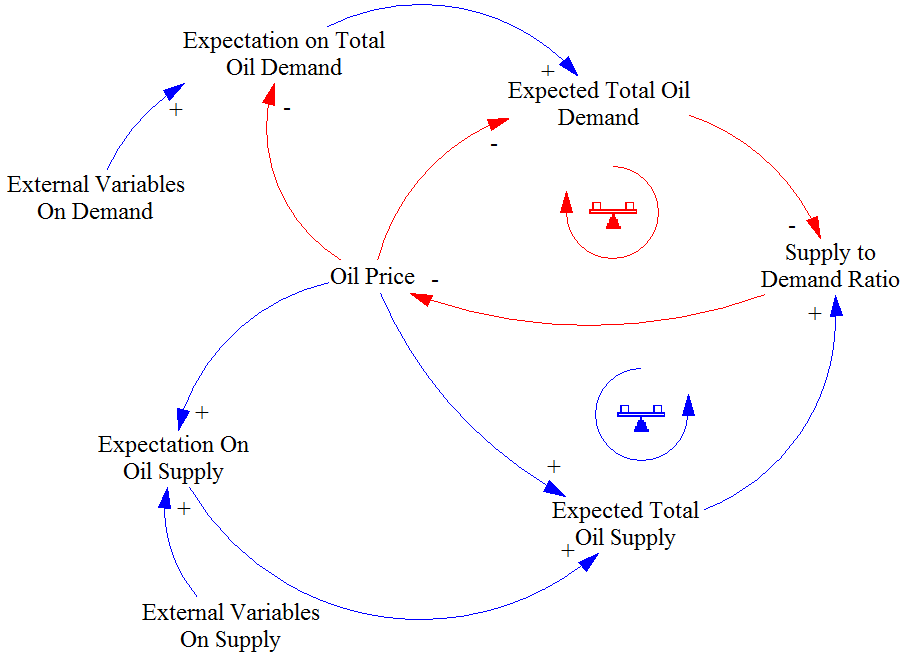}}
	\caption{The main loop and the sign of its arrows.\label{fig:Main Loop} }
\end{figure}

In this subsection, we provide the core structure of the model that involves the main variables in determining the oil price. Figure \ref{fig:Main Loop} depicts the main loop along with the sign of causal effects that the corresponding variables have on each other. The rest of this subsection explains the various parameters in this loop.

\textit{Components of the Main Loop:} Similar to \cite{rafieisakhaei2016modeling} the Main Loop consists of `Expected Total Oil Supply ($ TOS $)', `Expected Total Oil Demand ($ TOD $)', `Supply to Demand Ratio ($ TOS/TOD $)', `Expectation on Oil Supply ($ ExS $)' and `Expectation on Oil Demand ($ ExD $)' which interact with each other to determine the `Oil Price ($ P $)'. In addition, there are two variables named as `External Variables on Demand' and `External variables on Supply' which summarize external factors that may not directly influence the oil price (reflected in Fig. \ref{fig:Main Loop} as the External Variables on Supply and Demand, respectively). These factors can also depend on various variables that are determined in the economy as a whole (including the energy sector). We will elaborate them in the later sections.

\begin{table*}\centering
	\ra{1.3}\caption{The statistics of the data used for the regression in \eqref{eq:Supply to Demand Ratio and the Oil Price}.}\label{table:PCR-Ratio}
	\begin{tabular}[hb]{@{}c c c c c c@{}}\toprule
		& Total Demand (Mb/Quarter) & Total Supply (Mb/Quarter) & Supply to Demand Ratio & Price (USD) & Price Change Rate \\ \midrule
		Min	&	86.8	&	87.03	&	0.99	&	76.06	&	-12.53	\\
		Max	&	92.8	&	91.74	&	1.02	&	105.83	&	11.78	\\
		Mean	&	89.82	&	89.64	&	1	&	92.65	&	1.45	\\
		Median	&	89.7	&	89.71	&	1	&	94.04	&	1.19	\\
		STD	&	1.69	&	1.45	&	0.01	&	8.84	&	7.23	\\
		Kurtosis	&	-0.47	&	-1.11	&	0.28	&	-0.47	&	-0.74	\\
		Skewness	&	0.03	&	-0.21	&	0.61	&	-0.51	&	-0.46	\\
		\bottomrule
	\end{tabular}
\end{table*}

\textit{Supply to Demand Ratio and Oil Price:} As mentioned before, oil price like any other commodity in the economics, is mainly determined by the ratio of supply to demand \cite{fattouh2007drivers}. It is known that the price inversely affects the demand. Therefore, the higher the demand, the lower the supply to demand ratio and the higher the oil price. Moreover, increased oil price usually lowers the growth of the demand rate. On the other hand, oil price directly affects supply, which means that the higher the price, the higher the supply rate and the higher the supply to demand ratio. Thus, because of the inverse relation of the supply to demand ratio with the oil price, these loops stabilize and the price gets determined based on the ratio of the total supply and the total demand. We have used a first order regression using the West Texas Intermediate (WTI) index data in the normal states of the oil price to mathematically determine this negative feedback relationship, and this is what we discuss next.

\textit{Regression analysis for Price Change Rate (PCR) with the assumption of Euler method:} Now, given a time series real data of $ P(t),~ TOS(t) $, and $ TOD(t) $ at equidistant time steps $ t_k, 1\le k\le K $, we obtain the time series $ PCR(t_k) $ as follows:
\begin{align}
PCR(t_k) = \frac{P(t_{k+1})-P(t_{k})}{\delta t},
\end{align}
where $ \delta t := t_{k+1}-t_{k} $ is a constant period of time which we choose for our analysis and in fact represents the resolution of the data that we have used. Note that, $ \delta t $ in this analysis can be different than $ \Delta t = TIME ~STEP = 0.0625~ (Day) $ that we use in our simulations, however, $ \delta t/\Delta t\in \mathbb{N} $ where $ \mathbb{N} $ is the set of Natural numbers. Next, we do a linear regression analysis in ``R-Project'' to find the relation between $ PCR $' and $ TOS/TOD $ as follows:
\begin{align}\label{eq:Supply to Demand Ratio and the Oil Price}
PCR &= \alpha_1 TOS/TOD +\beta_1,
\end{align}
where $ \alpha_1 $ and $ \beta_1 $ are the results of the regression analysis on the quarterly data of supply, demand and price over the first quarter of 2010 to the second quarter of 2014 \cite{1_eia_gov_2015, 1_company_2013, 23_generator_2015} whose statistics is summarized in table \ref{table:PCR-Ratio}. The Stock and Flow model that is used for this part of the model is shown in Fig \ref{subfig:Price_model} which will be more clarified in the next parts.

Note that in obtaining equation \eqref{eq:Supply to Demand Ratio and the Oil Price}, we searched for a period of time during which the price volatility was not too high in the long-term analysis, and more importantly the period of time was recent enough to fit our model. The nature of the oil price imposes some limitations in doing the regression analysis. For instance, one cannot simply choose any arbitrary period of time and impose the regression results to fit for all the time. Because, in some periods, it is seen that the price does not follow a negative relation with the supply to demand ratio. This analysis confirms our previous conjecture that in the oil market, the oil price does not only depend on the supply to demand ratio. This is because, if this was the case, we should have seen a merely non-increasing relation between the ratio $ TOS/TOD $ and price change rate, which is not the case. Therefore, the period of time is important, as well. Moreover, as it is seen in the above equations, only a naive regression analysis is not enough to model the price and supply to demand ratio's relations. Since, we noticed that the residual remains high even with higher order polynomials. This shows that when we are considering long period of time, the relation is much more complex than some low-order polynomial. Moreover, we cannot just over-fit a very high order polynomial to the data, since, that would be over-simplification of the problem and would not be realistic or practical. Therefore, once again, this confirms that in the oil market, there are more hidden variables which heavily contribute to the price changes and should be taken into consideration in the models. As mentioned before, we call these variables the expectational variables and provide their related causal loops in the following sections. 

It is worth mentioning that given two known values of $ P(t_k) $ and $ P(t_{k+1}) $, since in the forth-order Runge-Kutta method four values are unknown, it is not possible to do the above regression method based on that method, unless some other heuristics and approximations are used. Therefore, we only use Euler's method for the regression analysis. However, in the integration, we use the Runge-Kutta method, as well.

\textit{Expectation on Oil Demand and Supply:} As mentioned before, unlike many other commodities, oil market is heavily influenced by the factors other than the total demand and supply \cite{chevillon2009physical}. Some of these variables like `Policies on Energy Consumption' that include policies on transportation are affected by the oil price whereas some of the others may not be directly influenced by the oil price. For instance, we have defined a variable named as `Economic Depression' which tries to model the effects of the economic depression that is still felt in the EU zone on the oil price. The `External Variables on Demand' and `External Variables on Supply' in Fig. \ref{fig:Main Loop} are dummy variables that reflect these parameters on the main loop.

\textit{Effects of the expectational variables on the Oil Price:} In our model, the only variable that directly affects the oil price, is Supply to Demand Ratio. The novelty of our work is in the way that we model the effects of the other factors on the oil price only through this ratio. As shown in Fig. \ref{fig:Main Loop}, the expectational variables act as a booster or suppressor of the total demand or supply. In other words, they only act as a coefficient that amplify the effects of the supply or demand on the price. Their corresponding values are determined internally in the model in their corresponding loops which will be explained later in the paper. The reason behind this modeling is that whenever there is a change in the $ ExD $, if it is a positive expectation, then the oil price increases. In such a case, it is as if the demand has slightly increased. On the other hand, if there is a negative expectation on the growth of the demand, the oil price faces a decrease. It is similar to the condition that the demand has decreased slightly. Even though the actual demand or supply might not change for some while, these expectations change the price much faster than the actual changes. As an instance, on the mid-August 2015, it was expected that China's demand will reduce, and this caused a very sharp decay in the oil price. In particular, the oil price went down from \$45.51 on 13th of Aug. to 4-year low record of \$38.18 on 25th of Aug. \cite{1_oilpricecom_2015}. It should be noted that, these expectations changes affect the price with the same sign that the demand affects the price. Therefore, we model all these variables as the expectational variables and aggregate their interacting effects on the `Expectation on the Oil Demand' variable, whose value is multiplied in the demand to amplify or suppress the effects of the actual demand on the oil price. The same line of reasoning applies for the supply side, as well. In a word, the expectational parameters model the expectational changes of the demand and supply, and therefore, have the same effect on the price that the demand and supply have. Hence, we have modeled them as if the total demand or supply have changed by a factor that is determined by them.

\section{Simulations and Scenarios}\label{sec:simulations}
In this section, we simulate several scenarios using Vensim PLE 6.3 \cite{VensimP} and bring their corresponding results to support our model. In each of the scenarios, some specific circumstance that can happen in the oil market have been investigated and the corresponding oil price changes have been reflected.

\begin{figure}[!t]
	\centering
	\includegraphics[width=3.5in,height=6cm]{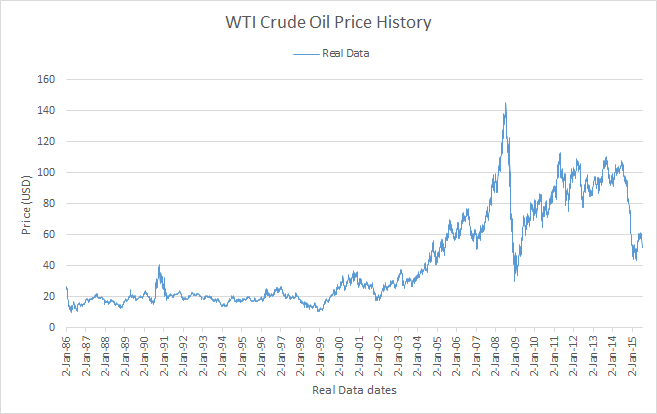}
	
	\caption{WTI oil price history.}
	\label{subfig:WTI_history}
\end{figure}
\begin{figure}[!t]
	\centering
	\includegraphics[width=3.5in,height=6cm]{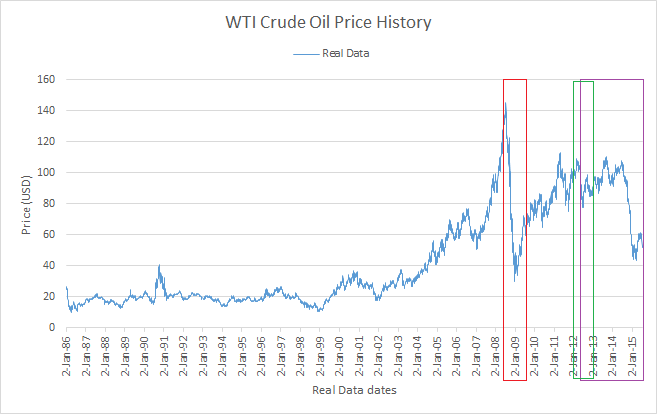}
	
	\caption{Parts of the WTI oil price history that we have used in our simulations, shown with rectangles.}
	\label{subfig:WTI_history_marked}
\end{figure}

\subsection{Scenario A: Neutral and Usual Case}
In this subsection, we simulate the model in a neutral condition, where there is no specific (or unexpected) event that can change the price significantly. The time horizon is considered to be a month and the initial oil price is supposed to be \$100.

\begin{figure}[!t]
	\centering
	\includegraphics[width=3.5in,height=6cm]{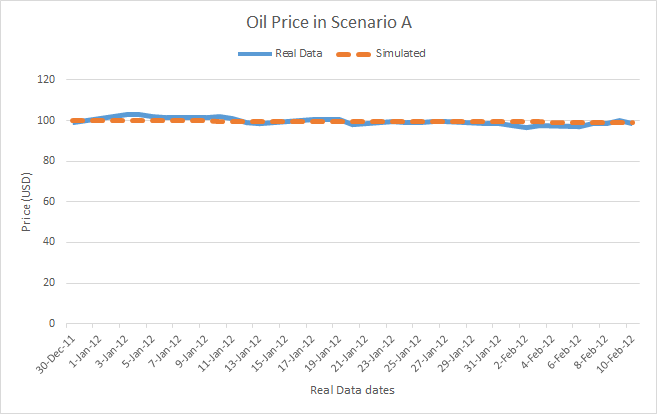}
	
	\caption{The Oil Price changes in Scenario A, outputs shown from Vensim and real data.}
	\label{subfig:Oil_Price_scenario_A_2}
\end{figure}

\textit{Results for the neutral case: }Figure \ref{subfig:Oil_Price_scenario_A_2} depicts the results corresponding to this case. As it is seen, the global oil price, is roughly constant over a time horizon of about one month, where the demand is predicted to be almost constant (since there is no significant event) and the supply is predicted to be slightly growing from US side. In our simulations, due to the slight amount of surplus in the supply, the price faces a very slow decay, reaching to \$99.16 within 43 days. The real data sequence that is used for this scenario involves the daily WTI price from 30th of Dec. 2011 to 10th of Jan. 2012.

\begin{figure}[!t]
	\centering
	
	\includegraphics[width=3.5in,height=6cm]{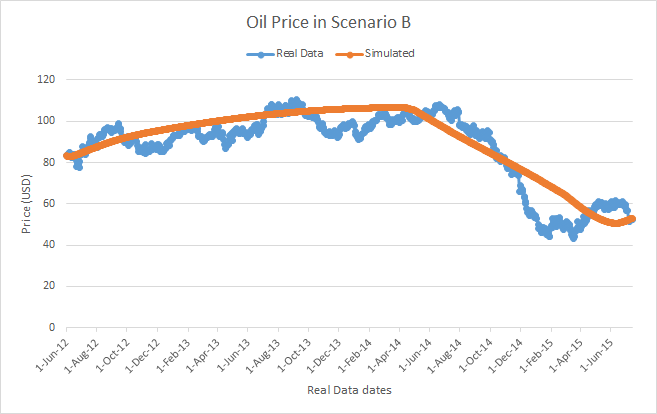}
	\caption{The Oil Price changes in Scenario B, outputs shown from Vensim and real data.} 
	\label{subfig:Scenario_B_vensim} 
	
\end{figure}

\begin{figure}[!t]
	\includegraphics[width=3.5in,height=6cm]{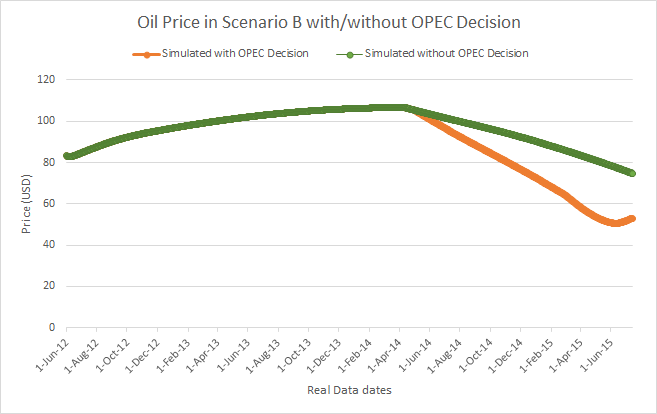}
	
	\caption{The Oil Price changes in Scenario B in a case that the OPEC Decision is set to zero (meaning that they allow the reduction of their production levels), outputs shown from Vensim and real data.}
	\label{subfig:Scenario_E_vensim}
\end{figure}

\subsection{Scenario B: Presence of Upset}
In this subsection, we consider a case where there is a Geopolitical Upset in the OPEC countries that results in the reduction of the OPEC's production. The results of this scenario is reflected in the Fig. \ref{subfig:Scenario_B_vensim}.

\textit{Effect of an upset on the Price:} In such a case, the immediate result is an increase on oil price due to the shortage in the supply which is boosted by the speculations on the future amount of supply. These speculations are reflected in the price by the factors Expectations on Supply. We additionally consider that the OPEC countries set-up a meeting with some delay in order to respond to the supply shortage. In such a case, two possible outcomes are possible. The first outcome is that they decide to keep the supply as it is, and let the prices go up. In the second scenario, they decide to use the spare supply resources to level up the supply to the previous levels. However, the resulting effect is that the prices remain high, although its change rate becomes nearly zero. This is due to the fact that the spare supply of the OPEC countries usually can be fed into the market with a delay and can remain at least for 90 days, but the speculations on the future of the oil productions still remain high due to the conflicts and the fact that the OPEC countries might not want to use their spare resources for a long period of time. In such a case, if the upsets are still existent, the price once again keeps climbing. In other word, despite the fact that the oil supply might change back to its normal amount, the expectations on oil supply still remain speculative. 

\section{Conclusion and Future Work}\label{sec:conclusion}

In this paper, we introduced a system dynamic model that explains the relations among some of the main variables in determining the global oil price. Through historic examples and analysis we showed that unlike many other commodities, oil price cannot be predicted only based on the aggregated supply and demand. Rather, it is through the `expected' demand and supply that the oil price gets determined. In our model we introduced the Expected Total Oil Demand and Expected Total Oil Supply to mathematically elaborate the effects of the expectations on the futures changes of supply and demand on the price. Particularly, we could still keep the traditional model of the commodity prices in which the price is obtained from the supply to demand ratio; however, instead of the actual values, we used the expected values of supply and demand in determining the ratio.

\bibliographystyle{IEEEtran}

\bibliography{MohammadRaf}

\end{document}